\documentclass[pra,twocolumn,showpacs,amsmath,amssymb,groupaddress]{revtex4-1} 
\usepackage{psfrag,graphicx}
\usepackage{amsfonts,amssymb,amsmath} 
\usepackage{graphicx}
\usepackage{dcolumn}
\usepackage{bm}
\usepackage{float}
\usepackage{hyperref}
\usepackage{accents}
\usepackage{bbding}
\usepackage{color,soul}
\usepackage{epstopdf} 
\newcommand{\ellipse}{\raisebox{-1pt}{\scalebox{1.3}[.4]{$\circ$}}}
\newcommand{\halo}{\accentset{\ellipse}}
\newcommand{\erf}[1]{Eq.~(\ref{#1})}
\newcommand{\beq}{\begin{equation}}
\newcommand{\eeq}{\end{equation}}
\newcommand{\nn}{\nonumber}

\newcommand{\erfs}[2]{Eqs.~(\ref{#1})--(\ref{#2})}

\newcommand{\sch}{Schr\"odinger}

\newcommand{\tp}{^{\top}}

\renewcommand{\c}{_{\text{C}}}
\newcommand{\ob}{_{\text{o}}}
\newcommand{\un}{_{\text{u}}}
\newcommand{\m}{_{\text{m}}}
\newcommand{\p}{_{\text{p}}}
\newcommand{\ex}[1]{\langle{#1}\rangle}
\newcommand{\dd}{{\rm d}}

\newcommand{\SHUR}{\sch-Heisenberg uncertainty relation}

\newcommand{\past}[1]{\overleftarrow{#1}}

\newcommand{\both}[1]{\overleftrightarrow{#1}}
\newcommand{\fil}{_{\text F}}
\newcommand{\rfil}{_{\text R}}
\newcommand{\sm}{_{\text S}}
\newcommand{\god}{_{\text T}}
\newcommand{\inv}{^{-1}}
\newcommand{\bx}{{\bf x}}
\newcommand{\bcx}{{\check{ \bf x}}}
\newcommand{\by}{{\bf y}}
\newcommand{\bv}{{\bf v}}
\newcommand{\bw}{{\bf w}}

\newcommand{\hV}{{\halo V}}

\newcommand{\xsm}{\ex{\hat \bx}\sm}
\newcommand{\xgod}{\ex{\hat \bx}\god}

\definecolor{nblue}{rgb}{0.06,0.3,0.73}
\definecolor{nblack}{rgb}{0,0,0}
\definecolor{nred}{rgb}{0.9,0.1,0.1}
\definecolor{nmagenta}{rgb}{0.7,0.0,0.3}

\newcommand{\red}{\color{nred}}

\newcommand{\blk}{\color{nblack}}

\newcommand{\hbx}{\hat\bx}

\newcommand{\hL}{{\halo \Lambda}}

\begin{document}
\title{Quantum State Smoothing for Linear Gaussian Systems}

\author{Kiarn T. Laverick}
\affiliation{Centre for Quantum Computation and Communication Technology 
(Australian Research Council), \\ Centre for Quantum Dynamics, Griffith University, Nathan, Queensland 
4111, Australia}
\author{Areeya Chantasri}
\affiliation{Centre for Quantum Computation and Communication Technology 
(Australian Research Council), \\ Centre for Quantum Dynamics, Griffith University, Nathan, Queensland 
4111, Australia}
\author{Howard M. Wiseman}
\affiliation{Centre for Quantum Computation and Communication Technology 
(Australian Research Council), \\ Centre for Quantum Dynamics, Griffith University, Nathan, Queensland 
4111, Australia}

\date{\today}
\begin{abstract}
Quantum state smoothing is a technique for assigning a valid quantum state to a partially observed 
dynamical system, using measurement records both prior and posterior to an estimation time. We show 
that the technique is greatly simplified for Linear 
Gaussian quantum systems, which have wide physical applicability. We derive a closed-form solution for 
the quantum smoothed state, which is more pure than the standard filtered state, whilst still being described by a physical quantum state, unlike other proposed quantum smoothing techniques. We 
apply the theory to an 
on-threshold optical parametric oscillator, 
exploring optimal conditions for purity recovery by smoothing. The role of quantum efficiency is elucidated, in both low and high efficiency limits.
\end{abstract}

\maketitle

Smoothing and filtering are techniques in classical estimation of dynamical systems to calculate 
probability density functions (PDFs) of quantities of interest at some time $t$, based on available data from noisy 
observation of 
such quantities in time. 
In filtering, the observed data up to time $t$ is used in the calculation. In smoothing, the
observed data both before (past) and after (future) $t$ can be used. For
dynamical systems where real-time estimation of the unknown parameters is not required, 
smoothing almost always gives more accurate 
estimates than filtering. In the quantum realm, numerous formalisms have been 
introduced which use past and future information~
\cite{ABL64,AAV88,Tsa09a,Tsa09b,GJM13,Cha13,Ohki15}. Many of these ideas have been applied, 
theoretically and experimentally, 
to the estimation of unknown classical parameters affecting quantum 
systems~\cite{WheatleyPRL10,TWC11,YonezawaSci12,Iwa13,Bud17,Hua18,LavWis18}, or of
hidden results of quantum 
measurements~\cite{HuletWV91,CamHua14,TanPRL15,RybMol15,Tan16,ZhaMol17}. The optimal 
improvement obtained by using future information 
in these applications comes from using classical Bayesian smoothing to obtain the PDF of the variables 
of interest. 

Despite such applications of smoothing to quantum parameter estimation, a quantum analogue 
for the classical smoothed state (i.e.~the PDF) was still missing. As quantum operators for a system at 
time $t$ do not 
commute with operators representing the results of later measurements on that system \cite{WisMil10}, a 
na\"ive generalisation of the classical smoothing technique would not result in a proper quantum state 
\cite{Tsa09b,GJM13,Ohki15}. 
As elucidated by Tsang \cite{Tsa09b} (see also the Supplemental Material of \cite{GJM13}), 
such a procedure would result in a  
``state'' that gives the (typically anomalous) weak-value \cite{AAV88} as its 
expectation value for any observable.
{\red A different way of doing quantum smoothing by na\"ive generalisation of the classical theory, also introduced by Tsang \cite{Tsa09b}, is to treat the Wigner distribution as if it were a classical PDF, giving rise to a smoothed Wigner distribution (SWD). The ``state'' corresponding to the SWD is also, in general, not a proper quantum state.}
In contrast to this,
Guevara and Wiseman \cite{GueWis15} recently proposed a theory of quantum state smoothing 
which also generalises classical smoothing but which gives a 
proper smoothed quantum state, 
i.e., both Hermitian and positive semi-definite.


The quantum state smoothing theory of Ref.~\cite{GueWis15} considers an open quantum system 
coupled to two baths (see Ref.~\cite{Bud17} for a similar idea). An observer,
Alice, monitors one bath and thereby obtains an ``observed''
measurement record ${\rm \bf O}$. Another observer, Bob (who is hidden from Alice), 
monitors the remaining 
bath, unobserved by Alice, and thereby obtains an 
``unobserved'' record ${\rm \bf U}$. If Alice knew $\past{\rm \bf U}$ as well as $\past{\rm \bf O}$ 
(the back-arrows indicating records in the past), she would have maximum knowledge of the quantum system, i.e.,~the 
``true'' state $\rho_{\past{\rm \bf O},\past{\rm \bf U}}$ at that time.
Thus, Alice's 
filtered and smoothed states can be defined in the same form of a conditioned state,
\beq\label{qsm}
\rho\c = \sum_{\past{\rm \bf U}} \wp\c(\past{\rm \bf U}) \rho_{\past{\rm \bf O},\past{\rm \bf U}}\,,
\eeq
where the summation is over all possible records unobserved by Alice.
For filtering ($\rho\c = \rho\fil$), the PDF of unobserved records is $\wp\c(\past{\rm \bf 
U}) = \wp(\past{\rm \bf U} | \past{\rm \bf O})$ conditioned on her {\em past record} $\past{\rm\bf O}$. 
For smoothing ($\rho\c = 
\rho\sm$), one has $\wp\c(\past{\rm \bf U}) = \wp(\past{\rm \bf U} | \both{\rm \bf O})$ 
conditioned on Alice's {\em past-future record} $\both{\rm\bf O}$. 
By construction, \erf{qsm} 
guarantees the positivity of the smoothed quantum state.

In this Letter we present the theory of quantum state smoothing for Linear Gaussian Quantum (LGQ)
systems. This can be applied to a large number of physical systems, e.g., multimodal light fields 
\cite{Bra05, Wee12}, optical and optomechanical systems \cite{WisDoh05, WisMil10,ZhaWie12,TsaNai12,BowTsa13, BowMil15,
Gen15, Wie15, VovRas17, ZhaMol17, Hua18, LiaJos18, OckDam18, SetTor18}, atomic ensembles \cite{Mad04,KohGer18,JimKol18}, and Bose-Einstein 
condensates 
\cite{Wad15}. 
Due to the nice properties of LGQ systems, we are able to obtain closed-form 
solutions for the smoothed LGQ state. 
This makes them much easier to study even than the two-level system originally considered in \cite{GueWis15}, as there is no need to generate numerically the numerous unobserved records appearing in the 
summation of \erf{qsm}. LQG smoothing only requires solving a few additional equations compared 
to classical smoothing for Linear Gaussian (LG) systems. The simplicity of our theory will enable 
easy application to numerous physical systems, and also 
allows analytical treatment of various measurement efficiency
regimes. We give such a treatment here for an 
optical parametric oscillator (OPO) on threshold \cite{WisDoh05, WisMil10}. 
As expected, our smoothed quantum state has higher purity than the usual 
filtered quantum state, while the \red SWD \blk state is often unphysical, with purity larger 
than one.

We begin by reviewing the necessary theoretical background of classical LG systems 
and LGQ systems. We then develop quantum state smoothing for LGQ systems and obtain analytic results in different limits. Finally, we apply LGQ smoothing to the on-threshold OPO.

{\em LG systems and classical smoothing.}--- 
Consider a classical dynamical system described by a vector of $M$ parameters $
\bx=\{x_1,x_2, ..., x_M\}\tp$. Here $\top$ denotes transpose. This system is regarded as an LG system if 
and only if it satisfies three conditions \cite{WisMil10,Hay01,Weinert01,vanTrees1,BroHwa12,Ein12,Fri12}. First, 
its evolution 
can be described by a linear Langevin equation
\beq\label{LLE}
{\rm d}\bx = A\bx {\rm d}t + E {\rm d}\bv\p\,.
\eeq
Here $A$ (the drift matrix) and $E$ are constant matrices and ${\rm d}\bv\p$ 
is the process noise, i.e., a vector of independent 
Wiener increments satisfying 
\beq \label{WeinCond}
{\rm E}[{\rm d}\bv\p] = {\bf 0}\,, \qquad {\rm d}\bv\p({\rm d}\bv\p)\tp = I {\rm d}t\,.
\eeq
Here ${\rm E}[...]$ represents an ensemble average, and 
$I$ is the $M\times M$ identity matrix. Second, knowledge about the system 
is conditioned on a measurement 
record $\by$ 
that is linear in $\bx$,
\beq\label{MeasCond}
\by{\rm d}t = C\bx{\rm d}t + {\rm d}\bv\m\,,
\eeq 
where $C$ is a constant matrix and the measurement noise ${\rm d}\bv\m$ is a vector of independent 
Wiener increments satisfying similar conditions to \erf{WeinCond}. It is possible for the process noise and 
the 
measurement noise to be correlated, e.g., from measurement back-action, which is described by a 
nonzero cross-correlation matrix $\Gamma$, computed from $\Gamma\tp {\rm d}t = E{\rm d}\bv\p({\rm d}
\bv\m)\tp$. The third condition is that the initial 
state of the system (i.e., the initial PDF of $\bx$, denoted as $\wp(\bx)|_{t=0}$) 
is Gaussian; then the linearity 
conditions (first and second) guarantee
the conditioned state will remain Gaussian:  
\beq
\wp\c(\bx) = g(\bx;\ex{\bx}\c,V\c)\,,
\eeq 
which is fully described by its mean $\ex{\bx}\c$ and 
variance (strictly, covariance matrix) $V\c \equiv \ex{\bx\bx\tp}\c - \ex{\bx}\c \ex{\bx}\c\tp$,  
throughout the entire evolution.

If the above criteria are met, one can compute a filtered LG state conditioned only on the 
past record (before the estimation time $t$). The filtered mean and variance are given by
,
\begin{align}
{\rm d}\ex{\bx}\fil=& \,\, A\ex{\bx}\fil {\rm d}t + {\cal K}^{+}[V\fil]{\rm d}\bw\fil\,,\label{fest}\\
\frac{{\rm d}V\fil}{{\rm d}t} = & \,\, AV\fil +V\fil A\tp + D - {\cal K}^{+} [V\fil] {\cal K}^{+} [V\fil]\tp\,,
\label{filvar}
\end{align}
where ${\rm d}\bw\fil \equiv \by{\rm d}t - C\ex{\bx}\fil{\rm d}t$ is a vector of innovations, $D = EE\tp$ is the 
diffusion matrix, and we have defined a ``kick" matrix, a function of $V$, via ${\cal K}^{\pm}[V] \equiv VC\tp 
\pm \Gamma\tp$. 
Initial conditions for these filtering equations are the mean and variance of 
the initial Gaussian state.

To solve for a smoothed LG state, one needs to include conditioning on the future record, 
which can be 
obtained from the retrofiltering equations 
\begin{align}
- {\rm d}\ex{\bx}\rfil =&  -A\ex{\bx}\rfil {\rm d}t + {\cal K}^{-} [V\rfil]{\rm d}\bw\rfil,\label{rest}\\
- \frac{{\rm d}V\rfil}{{\rm d}t} = & -AV\rfil - V\rfil A\tp + D - {\cal K}^{-} [V\rfil] {\cal K}^{-} [V\rfil]\tp\,,\label{rvar}
\end{align}
where ${\cal K}^-[V]$ was defined above and ${\rm d}\bw\rfil \equiv \by{\rm d}t - C\ex{\bx}\rfil{\rm d}t$. 
As the leading negative signs suggest, these equations are evolved backward in time, from a final 
condition at $t = T$. This is typically taken to be an uninformative 
PDF
. Combining the 
filtered and retrofiltered solutions \erfs{fest}{rvar}, one 
obtains a smoothed LG state conditioned on the 
entire measurement record~\cite{Hay01,Weinert01,BroHwa12,Ein12,vanTrees1},
\begin{align}\label{smest}
\ex{\bx}\sm &= V\sm(V\fil\inv\ex{\bx}\fil + V\rfil\inv\ex{\bx}\rfil)\,,\\
\label{csm}
V\sm &= (V\fil\inv + V\rfil\inv)\inv\,.
\end{align}

{\em LGQ systems.}---
For a quantum system analogous to the classical LG one, the system's 
observables require  
unbounded spectrums, represented 
by $N$ bosonic modes. We denote such a 
system by a vector of $M=2N$ observable
operators $\hbx = (\hat 
q_1,\hat p_1, ... ,\hat q_N,\hat p_N)\tp$, 
where $\hat q_{k}$ and $\hat p_{k}$ are canonically conjugate position and momentum operators for the 
$k$th mode, obeying the commutation relation $[\hat q_{k},\hat p_{ l}]=i\hbar \delta_{ kl}$. 
The system is called an LGQ system if its dynamical and measurement equations 
are isomorphic to those of a 
 classical LG system \cite{WisMil10,WisDoh05,Belav87,Bel92,DohJac99,Doh00}. For quantum 
systems there 
are additional constraints on the system's dynamics \cite{WisMil10}. For example the initial state must 
satisfy the \SHUR, $V+i\hbar\Sigma/2\geq0$. Here $\Sigma_{kl} = -i[\hat{x}_k,\hat{x}_l]$
is the symplectic matrix
and $V$ is the covariance matrix $V_{kl} = \ex{\hat x_{ k} \hat x_{ l} + \hat x_{ l} \hat 
x_{ k}}/2 - \ex{\hat x_{ k}}\ex{\hat x_{ l}}$, for ${\hat x}_{ k}$ being an element of $\hat \bx$ and $\ex{\cdot}$ being 
the usual quantum expectation value. 
These let us represent the quantum state 
of an LGQ system by its 
Gaussian Wigner function \cite{WisMil10}
defined as $W(\bcx) =  g(\bcx;\ex{\hat\bx},V)$, using dummy variable $\bcx$.

{\em Quantum state smoothing for LGQ systems.}--- 
We now apply the quantum state smoothing technique \cite{GueWis15} to LGQ systems. 
Following the Alice-Bob protocol introduced in \erf{qsm},
a true state of the LGQ system, denoted by the mean $\xgod$ and a variance $V\god$, is obtained given 
both $\past{\rm \bf O}$ and $\past{\rm \bf U}$ records. That is, 
the filtering equations 
\eqref{fest}-\eqref{filvar} apply, but conditioned both on Alice's observed record (of the form similar to 
\eqref{MeasCond})
\beq\label{obs}
\by\ob{\rm d}t = C\ob\xgod {\rm d}t + {\rm d}\bw\ob\,,
\eeq
and on Bob's record, unobserved by Alice,
$\by\un{\rm d}t = C\un\xgod{\rm d}t + {\rm d}\bw\un$,
with independent Wiener noises.  
The equations for the true state are
\begin{align}
{\rm d}\xgod =& \,\, A\xgod{\rm d}t + {\cal K}^{+}\ob[V\god]{\rm d}\bw\ob + {\cal K}^{+}\un[V\god]{\rm d}
\bw\un\,,\label{test}\\
\frac{{\rm d}V\god}{{\rm d}t} = & \,\, A V\god + V\god A\tp + D \nn\\
&- {\cal K}^{+}\ob[V\god]{\cal K}^{+} \ob[V\god]\tp - {\cal K}^{+}\un[V\god]{\cal K}^{+}\un[V\god]\tp\,, \label{truvar}
\end{align}
where ${\cal K}^\pm_{\rm r} [V] = 
VC\tp_{\rm r} + \Gamma\tp_{\rm r}$, for r $\in \{$o,u$\}$. 

Since Alice has no access to Bob's record, her conditioned state (filtered or smoothed) is 
obtained by summing over all 
possible true states 
of the 
system, with probability weights conditional on Alice's observed records ($\past{\rm \bf O}$ or $
\both{\rm \bf O}$, respectively) as in \erf{qsm}. 
For LGQ systems, the state depends on $\past{\rm \bf U}$ only via the mean, \erf{test}. 
Therefore, we can 
replace the (symbolic) sum in \erf{qsm} by an integral:
\beq\label{conv1}
\rho\c = \int \wp\c(\xgod) \rho\god(\ex{\hat\bx}\god) {\rm d}\xgod\,.
\eeq
Now let us define a ``haloed'' variable $\halo\bx = \ex{\hat\bx}\god$ 
for notational simplicity. We can 
replace the conditional state $\rho\c$ and true state $\rho\god$ with their Wigner functions. The latter is 
Gaussian: $g(\bcx;\halo\bx,V\god)$. The integral in \erf{conv1} convolves this with the PDF $
\wp\c(\halo\bx)$ conditioned on the observed records. 
This PDF is a conditioned (filtered or smoothed) LG distribution for 
$\halo\bx$, based on the observed data, $\wp\c(\halo\bx) = g(\halo\bx;\ex{\halo\bx}\c,\halo V\c)$, 
where  $\hV\c$ is the conditional variance for the variable $\halo\bx$ \cite{Supp}. 
As both 
functions inside the integral \erf{conv1} are Gaussian, the Wigner function for $\rho\c$ 
is also Gaussian:
\beq\label{conv}
g(\bcx;\ex{\hat\bx}\c,V\c) = \int g(\halo\bx;\ex{\halo\bx}\c,\halo V\c) g(\bcx;\halo\bx,V\god) {\rm d}
\halo\bx\,.
\eeq
By elementary properties of convolutions, we get the conditioned mean $\ex{\hat \bx}\c = \ex{\halo\bx}\c$ 
and the conditioned variance $V\c = \halo V\c + V\god$. This will allow us to solve for the 
filtered and smoothed quantum states for LGQ systems.

Now, all that remains is to apply classical LG 
estimation theory (filtering or smoothing) 
to determine $\ex{\halo\bx}\c$ and $\hV\c$.
We first obtain \cite{Supp} filtering equations for $\halo\bx$, using the past observed record \erf{obs},
\begin{align}
{\rm d}\ex{\halo\bx}\fil = &\,\, A\ex{\halo\bx}\fil{\rm d}t + {\cal K}^{+}\ob[\hV\fil+V\god] {\rm d}\halo\bw\fil\,,\\
\frac{{\rm d}\hV\fil}{{\rm d}t} = & \,\, A\hV\fil + \hV\fil A\tp + \halo D \nn\\
&- {\cal K}^{+}\ob[\hV\fil +V\god]{\cal K}^{+}
\ob[\hV\fil +V\god]\tp\,,
\end{align}
where we have defined 
$\halo D = \sum_{{\rm r} \in \{{\rm o},{\rm u}\}}{\cal K}^{+}_{\rm r}[V\god]{\cal K}^{+}_{\rm r}[V\god]\tp$,  
and ${\rm d}\halo\bw\fil = \by\ob{\rm d}t - C\ob\ex{\halo\bx}\fil {\rm d}t$. 
We also show in \cite{Supp} that this haloed filtered variance is related to 
the variance of the usual quantum filtered state $V\fil$ (computed without invoking the unobserved 
record) via $
V\fil = \hV\fil +V\god$ with the same mean $\ex{\hat \bx}\fil = \ex{\halo\bx}\fil$, consistent with the 
convolution \eqref{conv}. 
For the retrofiltering equations for $\halo\bx$, using the future record, we have
\begin{align}
-{\rm d}\ex{\halo\bx}\rfil = &-A\ex{\halo\bx}\rfil{\rm d}t + {\cal K}^{-}\ob[\hV\rfil - V\god]{\rm d}\halo\bw\rfil\,,\\
-\frac{{\rm d}\hV\rfil}{{\rm d}t} = & -A\hV\rfil - \hV\rfil A\tp + \halo D \nn\\
& - {\cal K}^{-}\ob[\hV\rfil -V\god]{\cal K}^{-}
\ob[\hV\rfil - V\god]\,,
\end{align}
which lead to a similar variance relation $V\rfil = \hV\rfil - V\god$ \cite{Supp}. However, the minus sign 
in the $\hV\rfil$ relation
indicates that the convolution \eqref{conv} does not apply for retrofiltering, which propagates in the 
backward direction in time.

We then combine the haloed filtering and retrofiltering equations, 
as in \erf{smest} and (\ref{csm}), to obtain the 
haloed smoothing equations, and using \eqref{conv}, we arrive at the LGQ state smoothing equations
\begin{align}
\xsm = (V\sm - V\god) [(V\fil &- V\god)\inv\ex{\halo \bx}\fil \nn\\
&+ (V\rfil+V\god)\inv \ex{\halo \bx}\rfil]\,,\label{estsm}\\
V\sm = \big[(V\fil - V\god )\inv &  + (V\rfil + V\god)\inv\big]\inv + V\god\,, \label{varsm}
\end{align}
as the main result of this Letter. In the classical limit, where there is no uncertainty relation for $V\god$ 
and we can let $V\god \to 0$, these reproduce classical LG smoothing, \erfs{smest}{csm}, as expected.

\begin{figure}[t!]
\includegraphics[scale=0.41]{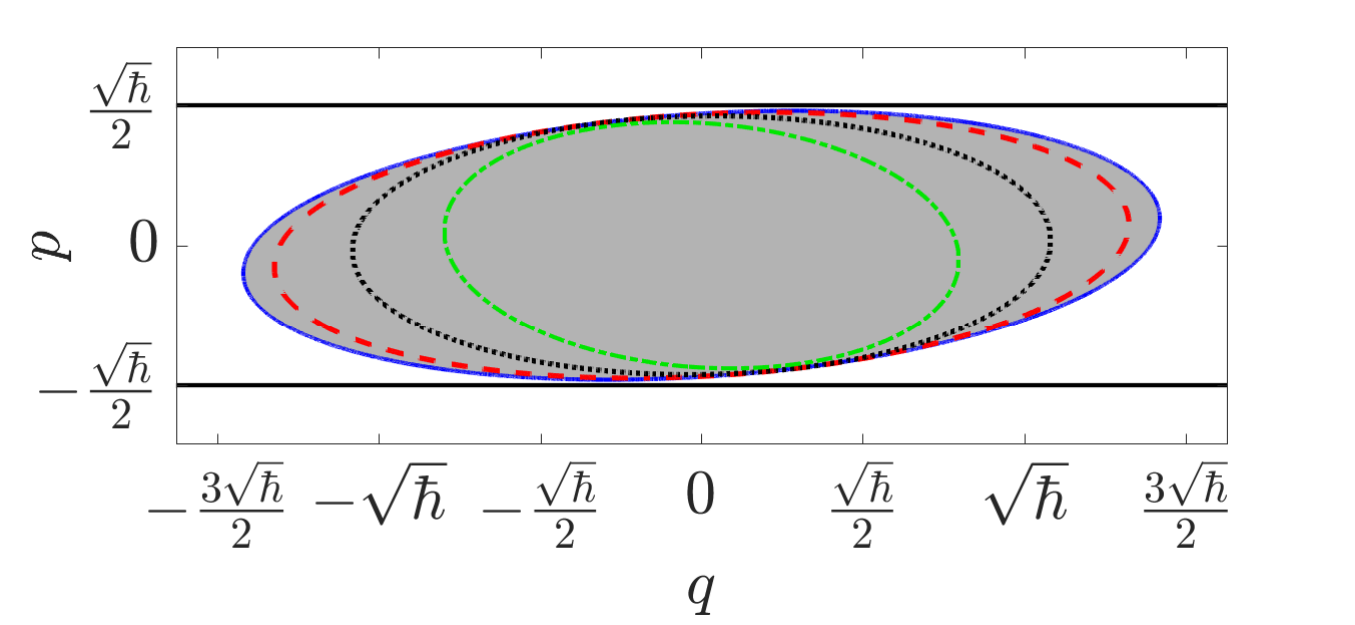}
\caption{(Colour online) Various long-time states of the on-threshold OPO system in \erf{eqopo}, 
represented by their 
{\red $e^{-1/2}$ contours, corresponding to 1-SD along the principle axes of the ellipse,}
in phase space, 
centred at the origin. The homodyne angles used by Alice and Bob ($\theta\ob$, $\theta\un$) are at the black dot in Fig.~\ref{RPR}.
The unconditional state (solid black) shows infinite 
and finite variances in $q$ and $p$, respectively, as a result of the damping and 
squeezing. Alice's filtered and smoothed states, are blue (filled grey) and 
dashed-red ellipses, respectively. The dotted-black ellipse shows the (pure) true state, conditioned on 
both Alice's and Bob's results, while the dot-dashed green ellipse shows the 
{\red SWD} ``state.''} 
\label{PSD}
\end{figure}

\begin{figure}[t]
\includegraphics[scale=0.321]{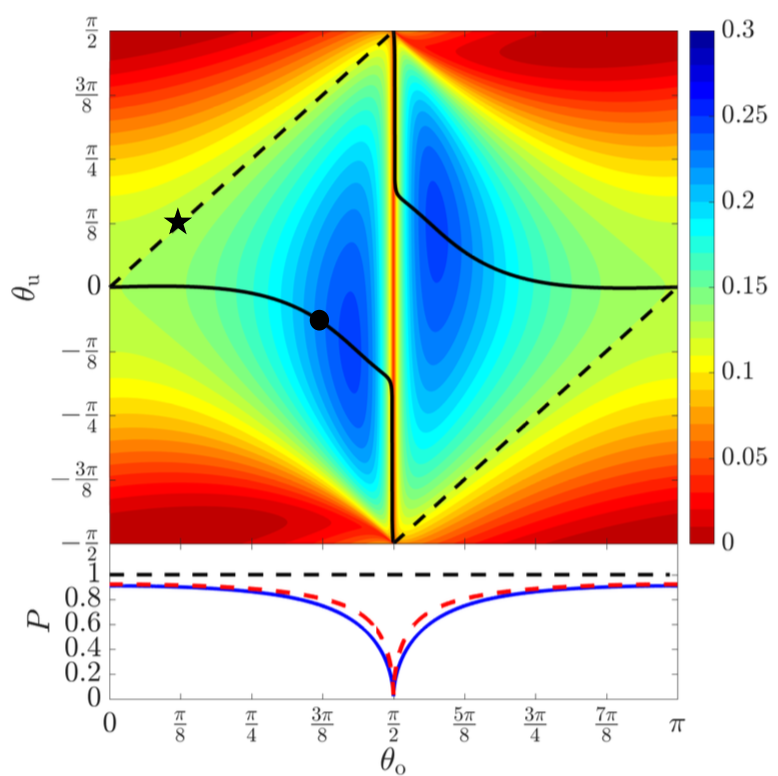}
\caption{(Colour online) (Top) Contour plots of the RPR, \erf{eqrpr}, for the 
OPO system for different values of observed 
and unobserved homodyne phases using $\eta\ob=0.5$. The dashed line represents $
\theta\ob=\theta\un$ and the solid line is 
the optimal $\theta\un$ (that giving the highest RPR for each value of $\theta\ob$). The circle and the star 
relate to Figs.~\ref{PSD} and \ref{Purity}, respectively. (Bottom) 
Purity for the OPO's filtered (solid blue) and smoothed (dashed red) states, 
choosing the optimal $\theta\un^{\rm opt}$ for each $\theta\ob$.}
\label{RPR}
\end{figure}

The advantages LGQ state smoothing offers over filtering are readily seen in Fig.~\ref{PSD}, where we 
note that the purity for a Gaussian state is defined as $P = (\hbar/2) \sqrt{|V|^{-1}}$ 
\cite{WisMil10} for a 
variance $V$. The 
smoothed state has a smaller variance (higher purity) than 
the filtered state, but has a larger variance than a pure state (purity less than unity).
In contrast, the {\red SWD} state for the same system (i.e., using \erfs{smest}{csm}) 
is unphysical (its ellipse is smaller than that of a pure state). 

Now that we have the closed-form expression for the smoothed LGQ state, we can investigate, 
in the steady state,  
some 
interesting limits in Alice's measurement efficiency $\eta\ob$, the fraction of the system output which is observed by Alice. 

If, as in the OPO system we will consider later, the unconditioned ($\eta\ob = 0$)  
variance diverges, then Alice's conditioned (filtered and retrofiltered) variances, if finite, must grow as 
$\eta\ob \to 0$. 
From \erfs{estsm}{varsm},
when $V\fil$ and $V\rfil$ are large, compared to $V\god$, the smoothed 
LGQ state reduces to the {\red SWD} state \erfs{smest}{csm}. The {\red SWD} state has the same form as classical smoothed states, which often have the same scaling as filtered states, but with a multiplicative constant improvement~\cite{TSL09,WheatleyPRL10,LavWis18}.  Consequently, in the limit $\eta\ob \to 0$, we expect 
 $P_{\red \rm SWD} =  P\sm \propto P\fil$ as functions of $\eta\ob$. 

In the opposite limit, $\eta\ob\to1$, we 
analytically show \cite{Supp} 
that the relative purity recovery 
(RPR),
\beq\label{eqrpr}
{\cal R} = \frac{P\sm - P\fil}{1- P\fil}\,,
\eeq
a measure of how much the purity is recovered from smoothing over filtering 
relative to the maximum recovery possible, usually scales with 
the unobserved efficiency. That is, ${\cal R} \propto \eta\un \equiv 1 - \eta\ob$.

{\em Example of the on-threshold OPO system.}--- 
We now apply quantum state smoothing to the on-threshold 
OPO \cite{WisMil10, WisDoh05}, an LGQ system with $N=1$ 
described by the master equation 
\beq\label{eqopo}
\hbar\dot\rho = -i[(\hat{q}\hat{p}+\hat{p}\hat{q})/2,\rho]+\mathcal{D}[\hat q+ i \hat{p}]\rho\,.
\eeq
The first term defines a Hamiltonian giving squeezing along the $p$-quadrature, while the second term 
describes the oscillator
damping. Here, the drift and diffusion matrices are $A={\rm diag}(0,-2)$ and $D=\hbar I$. 
Let us assume that Alice observes the damping channel via homodyne detection. 
Therefore, the matrix $C\ob$ in \eqref{obs} is 
$C\ob = 2\sqrt{\eta\ob/\hbar}\left(\cos\theta\ob,\sin\theta\ob\right)$,
where $\theta\ob$ is the homodyne phase~\cite{WisMil10, WisDoh05}. For simplicity, we 
assume Bob also performs a homodyne measurement, with a different phase $\theta\un$, 
so that
$C\un=2\sqrt{\eta\un/\hbar}\left( \cos\theta\un,\sin\theta\un\right)$. 
The measurement back-actions are 
described by matrices $\Gamma_{\rm r} = -\hbar C_{\rm r}/2$, for r $\in \{$o,u$\}$. 

We now solve for filtered and 
smoothed states for the OPO in steady state. We are particularly interested in the 
RPR \eqref{eqrpr} of smoothing over filtering, and in the combinations of 
homodyne phases that result in the largest RPR.
The RPR is always positive  (see Fig.~\ref{RPR}),  meaning that the smoothed quantum state 
always has higher purity than the corresponding filtered one. If Alice's phase $\theta\ob$ is 
fixed, one might guess that Bob's phase giving the best purity improvement should be the same, 
$\theta\un=\theta\ob$. However, that is not at all true (see Fig.~\ref{RPR}).  
The optimal $\theta\un^{\rm opt}$ is not a trivial function of 
$\theta\ob$. Rather, $\theta\un^{\rm opt} \approx 0$, i.e., Bob should measure the $q$-quadrature, which is presumably related to the fact that, without measurement in, the variance in $q$ diverges.



\begin{figure}[t]
\includegraphics[scale=0.23]{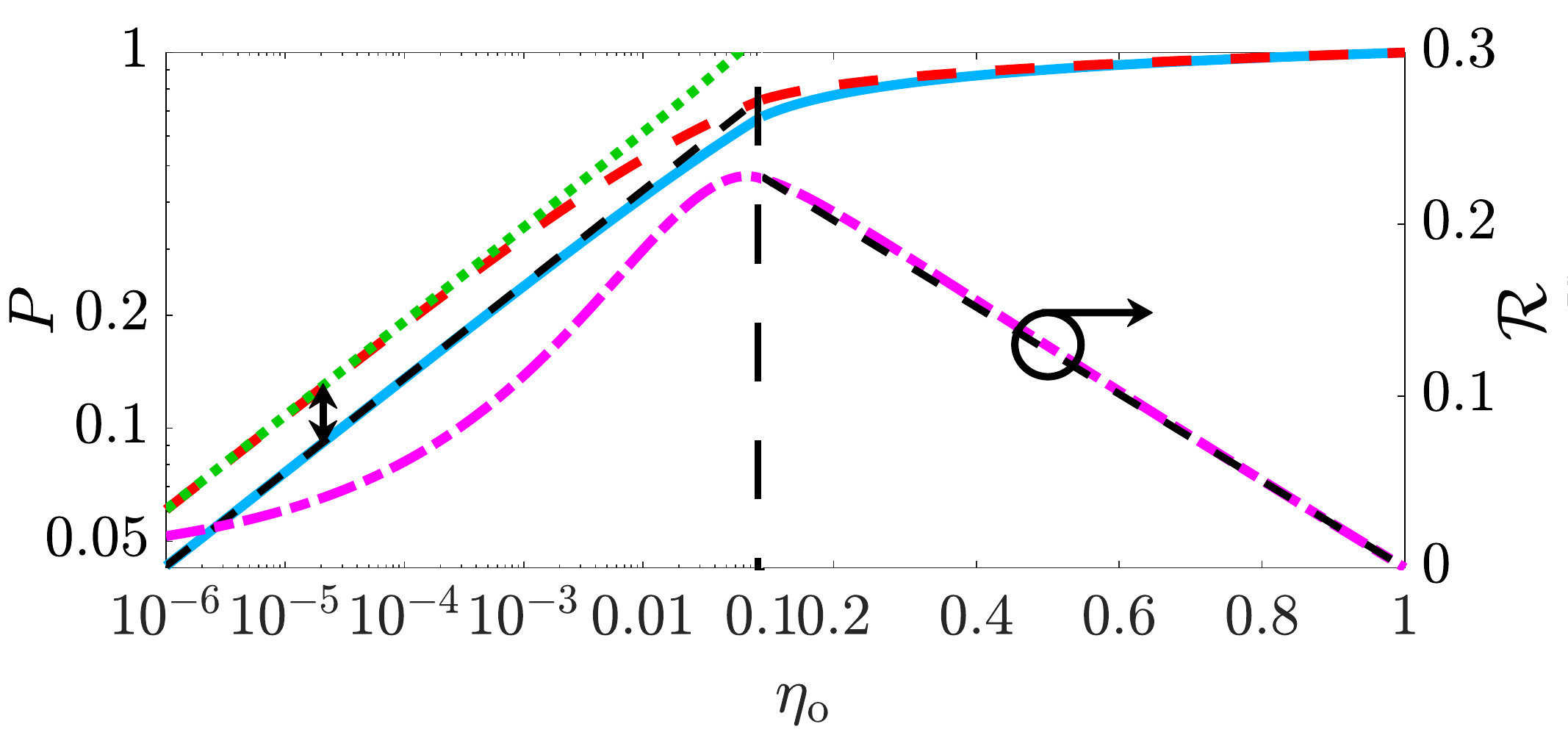}
\caption{(Colour online) Purities, and the RPR, \erf{eqrpr}, at the starred point in Fig.~\ref{RPR}, 
for the full range of Alice's measurement efficiency $\eta\ob$,
with the 
lower efficiencies plotted on a log scale and the higher efficiencies 
on a linear scale, where the dashed vertical line at $\eta\ob = 0.1$ indicates the split.
On both sides, we plot: purities of the filtered (solid blue), 
smoothed (dashed red) and the {\red SWD} (dotted green) states, all on a log scale (left-hand-side axis); 
and the RPR (${\cal R}$) (dot-dashed 
magenta), on a linear scale (the right-hand-side axis). For $\eta\ob \to 0$, 
$P\fil$ matches the simple analytic expression~\cite{Supp} $\sqrt{2|\cos\theta\ob|} \eta\ob^{1/4}$ (dashed black on left), 
 and smoothing gives a factor of $\sqrt{2}$ improvement~\cite{Supp}, as shown by the small $\updownarrow$ symbol. 
 For $\eta\ob \to 1$, the RPR is $\propto (1-\eta\ob)$ (dashed black on right). }
\label{Purity}
\end{figure}

We then examine, in Fig.~\ref{Purity}, the low and high efficiency limits for the OPO system at the starred point in Fig.~\ref{RPR}. 
As predicted 
earlier, in the limit $\eta\ob \to 0$ (left), the purities of the smoothed LGQ state and the {\red SWD} state are 
almost identical, and have a constant factor of improvement over that for 
filtering, as can be verified analytically~\cite{Supp}.
However, $P_{\red \rm SWD}$ begins to separate from $P\sm$ when the purities are no longer small, as 
the former proceeds to have purity greater than $1$ when $\eta\ob > 0.06$.
In the limit $\eta\ob \to 1$ (right), we see that the RPR has 
linear scaling in $\eta\un = 1-\eta\ob$, as expected. The approximation holds 
surprisingly well even when $\eta\un$ is not small. 

To conclude, we have developed the theory of quantum state smoothing, which gives valid smoothed 
quantum states, for LGQ systems, a class of systems with wide physical applicability.
By utilizing the Gaussian properties, we obtained closed-form smoothing solutions that do not 
require  
simulations of ensembles of
unobserved measurement records and corresponding true states. This enabled us to perform 
detailed 
analysis of the smoothed quantum state for various measurement regimes. 
A question for future work is to understand the
(numerically found) optimal strategy for greatest 
improvement in the purity. 
There are also interesting questions regarding how the smoothed LGQ variance 
(\ref{varsm}) would react to inserting an
invalid true state (i.e.,~one that does not solve \erf{truvar}). Finally, we could compare the smoothed LGQ 
state to other state estimation 
techniques using future information, such as the most likely 
path approach in Refs.~\cite{Cha13, Web14}. 

We acknowledge the traditional owners of the land on which this work was undertaken at Griffith University, the Yuggera people. This research is funded by the Australian Research Council Centre of Excellence Program CE170100012.  AC acknowledges the support of the Griffith University Postdoctoral Fellowship scheme.


%

\clearpage
\onecolumngrid
\appendix

\section{Haloed Filtering, Retrofiltering, and Smoothing}

We begin by deriving the haloed filtering and retrofiltering equations, and then show that $V\fil = \hV\fil 
+V\god$ and $V\rfil = \hV\rfil-V\god$. We start with the equations for the true state (Eq.~(13)-(14) in the main 
text) given both the observed and unobserved records,
\begin{align}
{\rm d}\xgod =& A\xgod{\rm d}t + {\cal K}^{+}\ob[V\god]{\rm d}\bw\ob + {\cal K}^{+}\un[V\god]{\rm d}
\bw\un\,,\label{lte}\\
\frac{{\rm d}V\god}{{\rm d}t} = A&V\god + V\god A\tp + D - {\cal K}^{+}\ob[V\god]{\cal K}^{+}
\ob[V\god]\tp - {\cal K}^{+}\un[V\god]{\cal K}^{+}\un[V\god]\tp\,. \label{VT}
\end{align}
We then use $\halo{\bx}=\ex{\hat\bx}\god$ and define $\halo{E}{\rm d}\halo{\bv}\p = {\cal K}^{+}\ob 
[V\god] {\rm d}\bw\ob +{\cal K}^{+}\un[V\god]{\rm d}\bw\un$, so that \erf{lte} is recast in a simple form as
\beq\label{hme}
{\rm d}\halo{\bx} = A\halo{\bx}{\rm d}t + \halo{E}{\rm d}\halo{\bv}\p\,.
\eeq
That is, \erf{hme} has the same form as the classical linear Langevin equation (Eq.~(2) in the main text) and 
$\halo\bx$ can be considered a classical parameter (a possible trajectory of the centroid of the true state). 
Moreover, Alice's observed measurement current (Eq.~(12) in the 
main text), 
\beq
\by\ob{\rm d}t  = C\ob\halo\bx{\rm d}t +{\rm d} \bw\ob\,,
\eeq
 has the same form as Eq.~(4) in the main text, with $\dd\bw\ob$ identified as $\dd\halo\bv\m$. The correlation of the measurement noise with the process noise is easily evaluated as 
\beq
\halo{\Gamma}^\top \dd t =  \halo E{\rm d}\halo\bv\p({\rm d}\halo\bv\m)\tp = {\cal K}^{+}\ob[V\god]{\rm d}t. 
\eeq
That is, $\halo{\Gamma} =  \Gamma _{\rm o} + C_{\rm o}V\god$. Similarly, we can define 
\beq
\halo{D} = \halo{E}\halo{E}\tp = {\cal K}^{+}\ob[V\god]{\cal K}^{+}\ob[V\god]\tp + {\cal K}^{+}\un[V\god]{\cal K}^{+}\un[V\god]\tp.
\eeq

\subsection{Filtering} 

From the above, we obtain the 
haloed filtering equations in a 
similar way using Eqs.~(6)--(7) in the main text, replacing $\bx$, $D$, $C$, and $\Gamma$ with $\halo\bx$, $\halo{D}$, $C\ob$, and $\halo\Gamma$, respectively.  The haloed filtering equations are then given by 
\begin{align}
&{\rm d}\ex{\halo\bx}\fil = A\ex{\halo\bx}\fil{\rm d}t + {\cal K}^{+}\ob[\hV\fil+V\god] {\rm d}\halo\bw\fil\,,\label{hfe}\\
&\frac{{\rm d}\hV\fil}{{\rm d}t} = A\hV\fil + \hV\fil A\tp + \halo D - {\cal K}^{+}\ob[\hV\fil +V\god]{\cal K}^{+}
\ob[\hV\fil +V\god]\tp\,,\label{hf}
\end{align}
as shown in Eqs.~(17)--(18) of the main text,
where ${\rm d}\halo\bw\fil\equiv \by\ob{\rm d}t - C\ob\ex{\halo\bx}\fil{\rm d}t$. 

To show the relation between the haloed filtering equation 
and the usual quantum filtering equations, we begin by 
recognising a different form of $\halo D$. From \erf{VT} we see that
\beq\label{hd}
\halo{D} = AV\god +V\god A\tp + D -\frac{{\rm d}V\god}{{\rm d}t}\,,
\eeq
and substituting this into \erf{hf} to get 
\beq
\frac{{\rm d}}{{\rm d}t}(\hV\fil + V\god) = A (\hV\fil +V\god) + (\hV\fil + V\god)A\tp + D - {\cal K}^{+}\ob[\hV\fil 
+V\god]{\cal K}^{+}\ob[\hV\fil +V\god]\,.
\eeq
If we use $V\fil = \hV\fil + V\god$ in the above equation, we obtain exactly Eq.~(7) in the main text, i.e., an 
equation for the filtered variance. Therefore, we can regard this variance $V\fil$ as the variance of the 
usual filtering for an LGQ system, which one could derive without the Alice-Bob protocol. We can also 
check that the haloed filtered mean is identical to the usual filtered mean
(Eq.~(6) of the main text, but for an LGQ system), as should be the case from the convolution. 
This can be seen by using $V\fil = \hV\fil + V\god$ in the haloed filtered estimate \erf{hfe}, which gives
\beq
{\rm d}\ex{\halo\bx}\fil = A\ex{\halo\bx}\fil{\rm d}t + {\cal K}^{+}\ob[V\fil]{\rm d}\halo\bw\fil\,.
\eeq
Considering the same initial conditions for $\ex{\halo\bx}\fil$ and $\ex{\hbx}\fil$, the haloed filtered 
mean will remain identical to the filtered mean, $\ex{\hat\bx}\fil = \ex{\halo\bx}\fil$, 
so the innovations will also be identical, with ${\rm d}\halo\bw\fil = {\rm d}\bw\fil  \equiv 
\by\ob\dd t - C\ob\ex{\hbx}\fil$.   Note 
that neither of these innovations is the same as the 
$\dd\bw\ob = \by\ob\dd t - C\ob\ex{\hbx}\god$ in \erf{lte}, as the latter is the innovation in Alice's record 
defined using the true state, i.e., from Bob's all-knowing point of view rather than Alice's.

\subsection{Retrofiltering and Smoothing}
Similarly to the filtering case, we can get the haloed retrofiltering equations using Eqs.~(8)--(9) in the main text
\begin{align}
-{\rm d}\ex{\halo\bx}\rfil &= -A\ex{\halo\bx}\rfil{\rm d}t + {\cal K}^{-}\ob[\hV\rfil - V\god]{\rm d}\halo\bw\rfil \,,\\
-\frac{{\rm d}\hV\rfil}{{\rm d}t} &= -A\hV\rfil - \hV\rfil A\tp + \halo D - {\cal K}^{-}\ob[\hV\rfil -V\god]{\cal K}^{-}
\ob[\hV\rfil - V\god]\,,\label{hrfil}
\end{align}
 where ${\rm d}\halo\bw\rfil \equiv \by\ob\dd t - C\ob\ex{\halo \bx}\rfil$.  
Now, adding \erf{hd} to \erf{hrfil}, we arrive at
\beq
-\frac{{\rm d}}{{\rm d}t}(\hV\rfil - V\god) = -A (\hV\rfil -V\god) - (\hV\rfil - V\god)A\tp + 
D - {\cal K}^{-}\ob[\hV\rfil -
V\god]{\cal K}^{-}\ob[\hV\rfil - V\god]\,,
\eeq
which is the equation for the retrofiltered variance $V\rfil$ and as a result we 
have $V\rfil = \hV\rfil - V\god$. 

The relation $V\rfil = \hV\rfil - V\god$ 
is interesting, as it shows the asymmetry between the filtered state and the retrofiltered effect. 
Note that $V_R$ is not the variance for a state conditioned only on the future observed record. Rather it is the variance of a POVM element for the future record. To obtain $V\rfil$ 
in general it will be easier 
to compute the inverse of $\hV\rfil$, rather than $\hV\rfil$ itself, for calculating the smoothed state. This is because the 
final condition on the retrofiltered
variance (haloed or not) is often taken to be infinite, as mentioned in the main text. 
Defining the inverse 
$\halo\Lambda\rfil = \hV\rfil^{-1}$, we use the relation $\frac{{\rm d}}{{\rm d}t}\left(\hV\rfil\hL\rfil\right) = 0$ to get
\beq
\frac{{\rm d}\hL\rfil}{{\rm d}t} = -\hL\rfil \frac{{\rm d}\hV\rfil}{{\rm d}t}\hL\rfil\,.
\eeq
From this we obtain   
\beq\label{LamR}
-\frac{{\rm d}\hL\rfil}{{\rm d}t} = \bar{A}\hL\rfil + \hL\rfil \bar{A}\tp - \hL\rfil\bar{D}\hL\rfil  + C\ob\tp C\ob\,,
\eeq
with $\bar{A} = A - \Gamma\ob C\ob - V\god C\ob\tp C\ob$ and 
$\bar{D} = \halo D - \Gamma\ob\tp\Gamma\ob - \Gamma\ob\tp C\ob V\god - V\god C\ob\tp\Gamma\ob 
- V\god C\ob\tp C\ob V\god$.
This way, the final condition is $\hL\rfil(T) = 0$ and the LGQ smoothed state in terms of $\hL\rfil$ and $V\fil$ is given by
\begin{align}
&\ex{\halo \bx}_{\rm S} = (V\sm - V\god) [(V\fil - V\god)\inv\ex{\halo \bx}\fil + \hL\rfil \ex{\halo \bx}\rfil]\,,\\
&V\sm = \left[(V\fil - V\god)\inv + \hL\rfil \right]\inv + V\god\,. \label{varsm-sm}
\end{align}

\section{Purities and RPR for different efficiency limits}
\subsection{Low Efficiency Limit}
For the low efficiency limit, specifically for the on-threshold OPO system, we will show here that the purity $P_{\rm C} \propto \eta\ob^{1/4}$, where ${\rm C} \in \{{\rm F},{\red \rm SWD}\}$.
We first consider the case $\eta\ob = 0$, i.e., no conditioning on measurement results, where the linear matrix equation for the steady state of the filtered variance $V\fil$ is given by
\beq\label{uncond}
AV\fil + V\fil A\tp + D = 0\,,
\eeq
with $A = {\rm diag}(0,-2)$ and $D = \hbar I$. Since the matrix $A$ for this case is not strictly stable (with one of its eigenvalues being zero), there is no stationary matrix solution for \eqref{uncond}, but in a long-time limit, we have~\cite{WisMil10}
\beq\label{fileta0}
V\fil \to \frac{\hbar}{2}\left[ \begin{array}{ccccc}
\infty & 0\\
0 & 1/2\\
\end{array}
\right] .
\eeq
We then consider the low efficiency case, for a small but non-zero observed efficiency $\eta_{\rm o} \rightarrow 0$. 
Now the filtered variance is given by Eq.~(7) in the main text. This equation leads to the variance in the $q$-quadrature (top-left element of the variance matrix) becoming finite, 
as long as the homodyne current contains some information about this quadrature ($\theta\ob \neq \pm\pi/2$). Explicitly, if we consider an arbitrary $V_F$ of the form
\beq
V\fil = \frac{\hbar}{2}\left[
\begin{array}{cc}
\alpha\fil&\beta\fil\\ 
\beta\fil&\gamma\fil\\
\end{array}\right]\,,
\eeq
we can obtain three relations for $\alpha\fil$, $\beta\fil$ and $\gamma\fil$, using Eq.~(7) (in the main text) and the matrices $C_{\rm o}$, $\Gamma_{\rm o}$ for the OPO system defined there: 
\begin{align}
1 &= \eta\ob\left[(\alpha\fil-1)\cos\theta\ob + \beta\fil\sin\theta\ob\right]^2, \label{loweffVF-1}\\
-\beta\fil & = \eta\ob \left((\alpha\fil-1)\cos\theta\ob + \beta\fil\sin\theta\ob\right)\left(\beta\fil\cos\theta\ob + (\gamma\fil-1)\sin\theta\ob\right),\\
1-2\gamma\fil & = \eta\ob \left[\beta\fil\cos\theta\ob + (\gamma\fil-1)\sin\theta\ob\right]^2 \label{loweffVF-3}.
\end{align}

To evaluate the purity of the filtered LGQ state, $P\fil$, in the low efficiency limit, we do not need to solve for the full solution of $V\fil$ from the above equations. The major contribution the measurement has to the variance is to bring the 
$q$-component, that is the top-left element, $\alpha\fil$, from infinity to a large but finite value; whereas the $\gamma\fil$ and $\beta\fil$ elements, describing variance in $p$-quadrature and covariance between the two quadratures, should still have values closed to those of the unconditional solution since they are finite even in the absence of any information, and the small amount of information in the low efficiency limit will make little difference. Consequently, we can treat $\alpha\fil$ as being much larger than $\gamma\fil$ and $\beta\fil$, where the former scales as an inverse order of $\eta\ob$ and the latter two are $O(\eta\ob^k)$ for $k \geq 0$. From this, we can only use \eqref{loweffVF-1} and solve for $\alpha\fil$ to leading
order in $\eta\ob$, giving
\beq
\alpha\fil \approx |\cos\theta\ob|\inv\eta\ob^{-1/2}\,.\\
\eeq
We now calculate the filtered purity using an assumption that the variance in $p$-quadrature represented by $\gamma\fil$ should still stay close to its unconditional value $1/2$, beginning with
\begin{align}
|V\fil| & \approx |2\cos\theta\ob|\inv\eta\ob^{-1/2} - \beta\fil^2 \approx |2\cos\theta\ob|\inv\eta\ob^{-1/2}\,.
\end{align}
Thus we get
\begin{align}
P\fil & = \frac{\hbar}{2}\sqrt{|V\fil|\inv} = \sqrt{2|\cos\theta\ob|}\, \eta\ob^{1/4}\label{ps}\,,
\end{align}
where we can see the $\eta\ob^{1/4}$ scaling. 

For the purity of the {\red smoothed Wigner distribution} 
state $P_{\red \rm SWD}$ in the small efficiency limit, we can use the intuition that the information used in the classical smoothing is twice as much in the filtering (considering the steady-state case), which should result in reducing the large variance in $q$-quadrature by half, i.e., $\alpha_{\rm \red SWD} \approx \frac{1}{2} \alpha\fil$, still much larger than $\beta_{\rm \red SWD}$ and $\gamma_{\rm \red SWD}$. As in the filtered case, we expect the latter two to remain little changed from their unconditioned values. Thus $|V_{\rm \red SWD}| \approx |4\cos\theta\ob|\inv\eta\ob^{-1/2}$ and 
\begin{align}
P_{\rm \red SWD} & = \frac{\hbar}{2}\sqrt{|V_{\rm \red SWD}\inv|} 
\approx 2\sqrt{|\cos\theta\ob|}\, \eta\ob^{1/4}\label{ps}\,,
\end{align}
where we see the constant factor of improvement $\sqrt{2}$ over filtering.

We now rigorously check our intuition by finding the exact solutions for both $P\fil$ and $P_{\rm \red SWD}$ in this limit, $\eta\ob \to 0$. Solving the full coupled equations \eqref{loweffVF-1}-\eqref{loweffVF-3}, we obtain $V\fil$ and $P\fil$ to leading orders in $\eta\ob$,
\beq
V\fil = \frac{\hbar}{2}\left[
\begin{array}{cc}
|\sec\theta\ob|\eta\ob^{-1/2}& \frac{1}{2} |\sin\theta\ob| \eta\ob^{1/2}\\ 
\frac{1}{2} |\sin\theta\ob| \eta\ob^{1/2} &1/2\\
\end{array}\right]\,, \qquad P\fil =\sqrt{2|\cos\theta\ob|}\, \eta\ob^{1/4}\,,
\eeq
as expected.
For the $V_{\rm \red SWD}$, we first need to calculate the 
retrofiltered variance $V\rfil$. We solve for the full solution of $V\rfil$ from Eq.~(9) in the main text (in the steady-state limit) to leading orders in $\eta\ob$,
\beq
V\rfil = \frac{\hbar}{2}\left[
\begin{array}{cc} 
|\sec\theta\ob|\eta\ob^{-1/2}& 2 |\csc\theta\ob| \eta\ob^{-1/2}   \\ 
2 |\csc\theta\ob| \eta\ob^{-1/2} & 2 |\csc \theta\ob|^2 \eta\ob^{-1}\\
\end{array}\right]\,.
\eeq 
Finally, we can calculate the {\red SWD} variance (using Eq.~(11) in the main text) and its purity, arriving at 
\beq
V_{\rm \red SWD} = \frac{\hbar}{2}\left[
\begin{array}{cc}
\frac{1}{2}|\sec\theta\ob|\eta\ob^{-1/2}& \frac{1}{2} |\sin\theta\ob| \eta\ob^{1/2}\\ 
\frac{1}{2} |\sin\theta\ob| \eta\ob^{1/2} &1/2\\
\end{array}\right]\,, \qquad P_{\rm \red SWD} =2\sqrt{|\cos\theta\ob|} \eta\ob^{1/4}\,,
\eeq
which are consistent with our intuitive approach.

\subsection{High Efficiency Limit}
In this section we derive the $\eta\un$ scaling for the steady-state relative purity recovery (RPR) in the high 
efficiency limit for Alice ($\eta\ob\to 1$), in general cases, not only for the OPO system. We first point out that with $\eta\ob=1$ (all available records are observed), the 
filtered and true variances are equal. So, if we 
express the filtered variance as $V\fil = V\god +Q$, 
it will typically be the case that $Q\to 0$ 
as $\eta\ob\to 1$, and we assume this to be so in all that follows. 
In this limit, we can see that the variance of the smoothed state, \erf{varsm-sm}, is
\begin{align}
V\sm &= \left[Q\inv + \hL\rfil\right]\inv + V\god\\
&= Q\left[1 + \hL\rfil Q\right]\inv + V\god\\
&\approx Q\left[1-\hL\rfil Q\right] + V\god\\
& = V\fil - Q\hL\rfil Q \label{q2}\,,
\end{align}
where the approximation holds since $Q$ is small. This nicely shows how information from the future, 
as expressed by $\hL\rfil \neq 0$, makes the smoothed variance smaller than the filtered one.

Now we show that the RPR (Eq.~(23) of the main text) scales as $O(Q)$.
The purity of a LGQ state is given by 
\beq
P_{\rm C} = \frac{\hbar}{2}\left[\sqrt{|V_{\rm C}|}\right]\inv\,,
\eeq
for ${\rm C} = {\rm F, S}$ or ${\rm T}$. The purity of the filtered state in the high efficiency limit is 
\begin{align}
P\fil &= \frac{\hbar}{2}\left[\sqrt{|V\god + Q|}\right]\inv,\\
& = \frac{\hbar}{2\sqrt{|V\god|}}\left[\sqrt{|I + V\god\inv Q|}\right]\inv,\\
& = P\god\left[\sqrt{|I + V\god\inv Q|}\right]\inv\,.
\end{align}
Now we need to evaluate $|I + Y|$, where $Y = V\god\inv Q$ 
is small.
Using the formula $|e^Y|= \exp[{\rm Tr}(Y)]$ \cite{Hall15}
and expanding 
the left and right exponential 
terms, we get, to leading order 
\beq
|I+Y| \approx 1+{\rm Tr}(Y)\,.
\eeq 
The purity of the filtered state is thus given by
\begin{align}
P\fil & \approx P\god\left[\sqrt{1 + {\rm Tr}(V\god\inv Q)}\right]\inv\\
& \approx P\god \left[1-{\rm Tr}\left(V\god\inv Q\right)/2\right]\,.
\end{align}

For the purity of the smoothed state, we express the smoothed variance as $V\sm = V\fil - X$, 
where $X = Q\halo\Lambda\rfil Q$. Following the similar derivation as for the purity of the 
filtered state, 
we obtain the purity of the smoothed state as $P\sm \approx P\fil\left[1 + {\rm Tr}\left(V\fil\inv Q\halo\Lambda\rfil Q\right)/2\right]$.
The RPR is then given by
\begin{align}
{\cal R} &=\frac{P\sm - P\fil}{1 - P\fil}\\
& = \frac{P\fil + P\fil{\rm Tr}\left(V\fil\inv Q\halo\Lambda\rfil Q\right)/2 - P\fil}{1-P\god + P\god{\rm Tr}\left(V\god\inv Q)\right)/2}\\
& = \frac{P\fil{\rm Tr}\left(V\fil\inv Q\halo\Lambda\rfil Q\right)/2}{1-P\god + P\god{\rm Tr}\left(V\god\inv Q)\right)/2}\,. 
\end{align}
If we consider that Bob observes the part unobserved by Alice's measurement, i.e.,~$\eta\un = 1-\eta\ob$, the true state will be a pure state ($P\god = 1$), as it is conditioned on all possible 
measurement records. The RPR then becomes 
\begin{align}
{\cal R} &\approx P\fil \frac{{\rm Tr}\left((V\god+Q)\inv Q\halo\Lambda\rfil Q\right)}{{\rm Tr}\left(V\god\inv Q\right)}\\
&\approx P\fil \frac{{\rm Tr}\left[\left(V\god\inv - V\god\inv QV\god\inv\right) Q\halo\Lambda\rfil Q\right]}{{\rm Tr}\left(V\god\inv Q\right)}\\ 
&\approx P\fil \frac{{\rm Tr}\left(V\god\inv Q\halo\Lambda\rfil Q\right)}{{\rm Tr}\left(V\god\inv Q\right)}\\
&= \frac{O(Q^2)}{O(Q)} \label{rprscale}= O(Q)\,.
\end{align}

Finally, all that is left is to check how $Q$ scales with the unobserved
measurement efficiency $\eta\un$.
Substituting in $V\fil = V\god + Q$ into Eq. (7) in the main text, we obtain
\beq
0 = A(V\god + Q) + (V\god + Q)A\tp + D -  {\cal K}^{+}\ob 
[V\god + Q] {\cal K}^{+}\ob [V\god + Q]\tp\,,
\eeq
considering the system to be in the steady state. Rearranging the above terms and using Eq.~(14)
in the main text (also in the steady state), we arrive at 
\beq\label{quad}
-\bar{A}Q - Q\bar{A}\tp +  Q{C}\ob\tp{C}\ob Q= \eta\un\bar{\cal K}^{+}\un[V\god]\bar{\cal K}^{+}
\un[V\god]\tp\,,
\eeq
where we are using $\bar{A}= A - \Gamma\ob\tp  C\ob - V\god {C}\tp\ob {C}\ob$  
as in Sec.~I~B above,  
and we have defined $ \bar{\cal K}^{+}_{\rm r}[V] = \sqrt{1/\eta\un}{\cal K}^{+}\un[V] $ 
so that in the limit $\eta\un\to 0$, all matrices in \erf{quad}, excluding $Q$, 
are independent of Bob's measurement efficiency $\eta\un$. That is because the matrices that 
are proportional to some positive power of Alice's efficiency $\eta\ob = 1-\eta\un$ 
have a limit independent of $\eta\un$ in the limit $\eta\un\to 0$. Now it might be thought that we can 
immediately discard the bilinear term in \erf{quad}, since $Q$ is small. This results in the linear equation 
\beq\label{QQ}
-\bar{A}Q - Q\bar{A}\tp = \eta\un\bar{\cal K}^{+}\un[V\god]\bar{\cal K}^{+}\un[V\god]\tp\,.
\eeq 
However this equation has a unique valid (positive semidefinite) solution for $Q$ if and only if $\bar{A}$ 
is Hurwitz. (A Hurwitz matrix is a real matrix where 
the real part of the eigenvalues are strictly negative.) Fortunately, we can expect this to be the case, for the following reason. 
In the limit $\eta\un\to 0$, $V\god = V\fil - Q \to V\fil$, and the matrix $\bar{A} \to M \equiv A - {\Gamma}\ob\tp C\ob - V\fil {C}\tp\ob {C}\ob$.
Now this matrix $M$ is well studied in control theory~\cite{WisMil10}; when the stationary filtered variance $V\fil$  
makes $M$ Hurwitz, it is said to be a stabilizing solution. There are well known conditions that ensure this to be the case~\cite{WisMil10} and these are satisfied for most systems of interest.  Moreover, these conditions are weaker for 
the case of quantum systems~\cite{WisMil10}.  Thus we will assume $\bar{A}$ 
to be Hurwitz. From \erf{QQ} we immediately see that $Q$, and consequently, from \erf{rprscale}, the relative purity recovery (${\cal R}$), scales as $\eta\un = 1-\eta\ob$ in the high efficiency limit $\eta\ob \to 1$. 
We can see this scaling explicitly in the 2-dimensional case, relevant to the OPO system,  where the solution to the linear matrix equation \erf{QQ} is \cite{WallsMilb94,WisMil10}
\beq
Q = \eta\un \frac{|\bar{A}| \bar{\cal K}^{+}\un[V\god]\bar{\cal K}^{+}\un[V\god]\tp+ (\bar{A} - I {\rm Tr}[\bar{A}]) \bar{\cal K}^{+}\un[V\god]\bar{\cal K}^{+}\un[V\god]\tp(\bar{A} - I{\rm Tr}[\bar{A}])\tp}{2{\rm Tr}[\bar{A}]|\bar{A}|}\,.
\eeq

\end{document}